\documentclass{sig-alternate-per}

\usepackage[margin=1in]{geometry}

\usepackage{amsmath}
\usepackage{framed}
\usepackage{url}

\begin{document}

\title{Hierarchical Analyses Applied to Computer System Performance: 
Review and Call for Further Studies}

\date{\vspace{-5ex}}
\author{
\textit{Alexander Thomasian}           \\
\textit{Thomasian and Associates}      \\
\textit{Pleasantville, NY}             \\
\textit{alexthomasian@gmail.com}       
}
\date{}
\maketitle

\begin{abstract}
We review studies based on analytic (A) and simulation (S) methods 
for hierarchical performance analysis of Queueing Network - QN models.
A at lower and S at higher level have been applied most.  
The proposed methods result in an order of magnitude reduction 
in performance evaluation cost with respect to simulation.
The computational cost at the lower level to obtain an exact solution 
is reduced when the computer system can be modeled as a product-form QN amenable to a low cost solution.
A Continuous Time Markov Chain - CTMC or discrete-event simulation can then be used at the higher level.
We first consider a multiprogrammed transaction - txn processing system 
with Poisson arrivals and predeclared lock requests.
Txns with lock conflicts with active txns are held in a FCFS queue
and txns are activated after they acquire all requested locks.
Txn throughputs obtained by the analysis of multiprogrammed computer systems 
serve as the transition rates in a higher level CTMC to determine txn response times.
We next analyze a task system where task precedence relationships 
are specified by a directed acyclic graph to determine its makespan. 
Task service demands are specified on the devices of a computer system. 
The composition of tasks in execution determines their processing time and throughputs,
which serve as transition rates among the states of the CTMC model.
To reduce memory space requirements the CTMC is built and solved one set of task completions at a time. 
As a third example we consider the hierarchical simulation 
of a timesharing system with two user classes.
Txn throughputs in processing various combinations of requests 
are obtained by analyzing a closed product-form QN model.
A discrete event simulator is provided.
More detailed QN modeling parameters, such as the distribution of the 
number of cycles of tasks consisting of Fork/Join (F/J) requests affect performance. 
This detail can be taken into account in Schwetman's hybrid simulation method, 
which counts remaining number of cycles in CSM-like queueing model.
We discuss an extension to hybrid simulation  
to adjust job service demands according to elapsed time, rather than counting cycles.
A section reviewing related studies is provided.
Equilibrium Point Analysis to reduce the computational cost 
in applying hierarchical analysis is presented in the Appendix.
The discussion is applicable to performance modeling of manufacturing systems.
\end{abstract}

%--------------------------------------------------
\section{Introduction}
\vspace{5mm}

Product-form {\it Queueing Networks - QNs} were initially restricted 
to single- and multi-server nodes with exponential service times and FCFS scheduling Jackson 1957 \cite{Jack57}.
Product-form QN's were extended to {\it Processor-Sharing - PS} 
and {\it Last-Come First-Served Preemptive Resume - LCFSPR} Kleinrock 1976 \cite{Klei76} and delay servers.
The latter servers allow general service times according to the BCMP theorem Baskett et al. 1975 \cite{BCMP75}. 
PS is an extreme form of round-robin CPU scheduling, 
where each job is allowed a quantum $q \rightarrow 0$ time units before preemption \cite{Klei76}. 

The {\it Buzen Convolution Algorithm - BCA} Buzen 1973 \cite{Buze73}
was a first step in efficiently solving product-form closed QNs,
where completed jobs are immediately replaced by a new job.
BCA was applied to the {\it Central server Model - CSM} described below.

%%%%%%%%%%%%%%%%%%%%%%%%%%%%%%%%%%%%%%%%%%%%%%%%%%%%%%%%%%%%%%%%%%%%%%%%%%%%%%%%5
\begin{framed}
\subsubsection*{Central Server Model - CSM}

CSM is a closed QN model of a multiprogrammed computer system Buzen 1973 \cite{Buze73},
which consists of a CPU and multiple disks.
Jobs alternate between CPU and disk processing until they are completed.
Completed jobs are immediately replaced by another job in closed systems 
or after think times modeled as a delay servers in time-sharing systems.

The CPU is designated as the central station ${\cal S}_1$
and the $N-1$ disks as peripheral stations ${\cal S}_n, 2 \leq n \leq N$.
Given the state transition probabilities ${\cal S}_i \xrightarrow{p_{i,j}} {\cal S}_j$
the following transitions are applicable to CSM.
$$p_{1,n}, \hspace{2mm}2 \leq n \leq N, \hspace{3mm}p_{n,1}=1, \hspace{2mm} 2 \leq n \leq N$$

The self-transition $p_{1,1}=1 - \sum_{n=2}^N p_{1,n}$ implies the completion of a job
in a closed QN (or a job that leaves the system in an open QN).
The number of visits to the CPU ($\bar{v}_1$) is given 
by the geometric distribution Trivedi 2001 \cite{Triv01}.
$$q_k = p_{1,1} (1-p_{1,1})^{k-1}, k \geq 1 \hspace{2mm}\bar{k}=v_1 = 1/p_{1,1}.$$

The relative number of visits to the stations is obtained by solving
$$\underbar{v}=\underbar{v}{\bf P}.$$
It follows $v_n = p_{1,n} v_1 = p_{1,n}/p(1,1), 2 \leq n \leq N$.

Given mean service time at ${\cal S}_n$ per visit is $\bar{x}_n$,
the mean loading per job is $X_n=v_n \bar{x}_n, 1 \leq n \leq N$.

Sauer and Chandy 1975 \cite{SaCh75} consider the analysis of a CSM
with a  CPU with  FCFS and priority (nonpreemptive and preemptive) scheduling 
and nonexponential service time scheduling.

\end{framed}
%%%%%%%%%%%%%%%%%%%%%%%%%%%%%%%%%%%%%%%%%%%%%%%%%%%%%%%%%%%%%%%%%%%%%%%%%
 
CSM is a single job class QN but BCA was extended to multiple job classes 
by Buzen and coworkers at BGS for inclusion in the BEST/1 capacity planning tool for MVS OS, renamed z/OS \cite{Buz+78}.
This extension was also done independently at IBM by Reiser and Kobayashi in 1975 \cite{ReKo75}.
The tutorial by Williams and Bhandiwad 1976 \cite{WiBh76} on the use of generating functions
in developing the convolution algorithm for multiple job classes 
was extended in Thomasian and Nadji 1981 \cite{ThNa81}.

{\it Mean Value Analysis - MVA} method developed by Reiser and Lavenberg 1980 \cite{ReLa80,Reis00}
has the same computational cost as BCA, but higher memory requirements.
It has numerical problems in dealing with state-dependent servers 
whose service rate varies with the number of jobs, such as multiserver queues.
MVA on the other hand has led to several low-cost, iterative solution methods, such as Bard-Schweitzer, 
see e.g., Lazowska et al. 1984 \cite{LZGS84}, 
and Linearizer Chandy and Neuse 1982 \cite{ChNe82}
Efficient approximate computational methods were later developed 
by extending MVA to non-product-form QNs, 
such as FCFS scheduling with general service times Lazowska et al. 1984 \cite{LZGS84}.

Analysis of open (resp. closed) QNs requires the arrival rate
(resp. degree of concurrency or {\it MultiProgramming Level - MPL}) and service demands or loadings. 
which are the product of the mean number of job visits to the devices of a computer  
and the mean service time per visit \cite{Buze73}.

IBM's {\it Software Measurement Facility - SMF} measures 
the mean time computer devices (CPU and disk) are busy serving tasks.
The service demands differ according to job class, e.g., batch versus online transactions - txns.
Given the {\it MultiProgramming Level - MPL}
and service demands BEST/1 Buzen et al. 1978 \cite{Buz+78,Buze78} 
and MAP Lazowska et al. 1984 \cite{LZGS84} capacity planning tools 
use QN analysis to obtained performance metrics of interest such as 
job throughput, device utilizations, response times, and queuelengths 
\cite{Lave83,LZGS84,BGdT06,KoMa09}. 

When tasks are to be processed at heterogeneous computer systems,
e.g., with different CPU speeds or different storage systems: 
{\it Hard Disk Drives - HDDs} versus {\it Solid State Disks - SSDs}.
task processing requirements should be specified in device independent manner,
e.g., program pathlenghts which can be converted to CPU time based on its MIPS.
%A similar argument applies to data stored on magnetic {\it Hard Disk Drives - HDDs} with different speeds.
%Orders of magnitude lower latency and higher bandwidth can 
%be attained when data is stored on {\it Solid State Disks - SSDs}.

%%%%%%%%%%%%%%%%%%%%%%%%%%%%%%%%%%%%%%%%%%%%%%%%%%%%%%%%%%%%%%%
\subsection*{Processing Time of Fork/Join Requests}

As an example of hierarchical modeling consider the time 
it takes to execute the tasks of a single k-way F/J request on a multiprogrammed computer system.
The approximate hierarchical analysis method based on decomposition Courtois 1975 \cite{Cour75}
(see e.g., Section 9.3.1 in Lazowska et al. 1984 \cite{LZGS84})
uses a {\it Flow-Equivalent Service Center - FESC},
whose throughput characteristic is obtained by analyzing the underlying QN model. 

The $K$ tasks are assumed to have identical service demands 
that can be activated concurrently by a computer system with maximum MPL $M_{max} \geq K$,
Task completion rates can be determined at low cost yielding $T(k), 1 \leq k \leq K$.
In hierarchical modeling task completions are assumed to be exponentially distributed 
and the completing time of $K$ tasks van be determined by a death process \cite{Klei75}. 
$$S_k \stackrel{T(k)} \longrightarrow S_{k-1}, K \geq k \geq 1.$$
The completion time of F/J requests is:
$$R_{F/J}(K) = \sum_{k=1}^K R(k)\mbox{  where  }R(k) = [T(k)]^{-1}.$$

%%%%%%%%%%%%%%%%%%%%%%%%%%%%%%%%%%%%%%%%%%%%%%%%%%%%%%%%%%%%%%%
\begin{framed}
\subsection{Degree of Concurrency Constraints}
\vspace{5mm}

Product form QN models of computer systems with Poisson arrivals 
with rate $\lambda$ is not amenable to a direct solution   
when the degree of concurrency or MPL, say $M_{max}$, is taken into account, 
because the number of jobs at the QN may exceed $M_{max}$.  

Assuming that the throughout characteristic $T(M)$ is a nondecreasing function of $M$
the maximum system throughput $\lambda_{max} < T(M_{max})$,
since otherwise the system will become saturated, 
i.e., a queue of infinite length will be formed Kleinrock 1975 \cite{Klei75}.

The MPL constraint is taken into account in applying a birth-death queueing model
with arrival rate $\lambda$ and service rate $\mu_k$ by ``flattening'' the
throughput characteristic beyond $T(M_{max})$ for the FESC as given by Eq. (\ref{eq:flat}):

\vspace{-1mm}
\begin{eqnarray}\label{eq:flat}
\mu_k = 
\begin{cases}
T(k), \hspace{2mm} 1 \leq k \leq M_{max}                     \\
T(M_{max}),\hspace{7mm}  k \geq M_{max}
\end{cases}
\end{eqnarray}

The mean number of tasks in the system and the memory queue (MemQ) are obtained as follows.

\begin{eqnarray}
\bar{N}  = \sum_{k \geq 1} k p_k, \hspace{2mm} \bar{N}_{MemQ} =\sum_{k \geq M_{max}} (k - M_{max}) p_k.
\end{eqnarray}

The mean response time in the systems and mean waiting time in the queue are obtained 
by applying Little's result by dividing by the task arrival rate $\lambda$ \cite{Klei75}: 
$$R_{system}=N_{systems}/\lambda\mbox{ and }W_{memQ}=\bar{N}_{MemQ}/\lambda.$$

State probabilities of the birth-death process with arrival rate $\lambda$ and
processing rate $\mu_k, k\geq 1$ are obtained  by setting $S=p_0=1$, $\bar{N}=0$ 
\begin{eqnarray}
p_k = (\lambda/\mu_k) p_{k-1}, \hspace{2mm} S += p_k , \hspace{2mm} \bar{N} += k p_k \\
\nonumber
k \geq 1 \mbox{ till } p_k \leq \epsilon , \bar{N} = \bar{N}/S, R=\bar{N}/\lambda.  
\end{eqnarray}

\end{framed}
%%%%%%%%%%%%%%%%%%%%%%%%%%%%%%%%%%%%%%%%%%%%%%%%%%%%%%%%%%%%%%%%%%

{\bf Example I: Transactions with predeclared lock requests:}
Txns with predeclared lock requests arriving according 
to a Poisson process with frequency $f_j, 1 \leq j \leq J$
can execute concurrently if they have no conflicts Thomasian 1985 \cite{Thom85}.
Txn response time is the sum of the queueing delay in a FCFS queue 
awaiting the acquisition of all locks at which point the txn is activated 
and task execution time at the computer system.
It is assumed that the maximum MPL is not a constraint.

An approximate solution is also presented by analyzing the QN for various degrees of concurrency 
and using the resulting throughout as if there is a single job class.

%-----------------------------------------------------------------------------
{\bf Example II: Tasks with precedence relationships:}  
Task precedence relationships are specified by a {\it directed acyclic graph - dag}
which is referred to as a task system in Coffman and Denning 1973 \cite{CoDe73}.
Task processing times are specified by their execution time on the devices of a single computer.
An optimal scheduling algorithm for two processors is presented in this book,
while scheduling with more processors is explored in Adam et al. 1973 \cite{AdCD74}.
The results were compared against the bound by Fernandez and Bussell 1973 \cite{FeBu73}.

A {\it Continuous Time Markov Chain - CTMC} at the higher level model 
and a product-form QN model for task execution on a multiprogrammed computer system
is considered in Thomasian and Bay 1986 \cite{ThBa86} 

%--------------------------------------------------------------------
{\bf Example III: Timesharing system:}
Simulation is a flexible approach for the higher level and its use is illustrated 
in the context of a timesharing system with two job classes Sauer 1981 \cite{Saue81}.
At the lower modeling level task throughputs are obtained by analyzing product-form closed QN model. 
Section \ref{sec:timesharing} specifies a discrete event simulation 
for higher level analysis of a timesharing system, 
whose tasks are processed in a multiprogrammed computer system. 

%---------------------------------------------------------------
{\bf Example IV: Fork/Join Analysis:}
A detailed QN model is required in evaluating the performance 
of a {\it Fork/Join - F/J} systems Thomasian 2014 \cite{Thom14}.
This is because the completion time of several tasks started concurrently 
is affected by the distribution of the number of processing cycles.
Detailed modeling can be better handled by hybrid simulation Schwetman 1978 \cite{Schw78}.

The paper is organized as follows.
Section \ref{sec:static} discusses a hierarchical model 
for analyzing a txn processing system with predeclared lock requests.
Section \ref{sec:tasksystem} determines the makespan of a task system,
whose tasks execute at a computer system. 
Section \ref{sec:timesharing}  describes a simulation model
to estimate the mean response times of timesharing requests.
The effect of transition probabilities on completion times is discussed in section \ref{sec:distr}.
Section \ref{sec:hybrid} describes the hybrid simulation method 
and propose extensions to it which were earlier discussed on \cite{ThBa83}, 
which requires further investigation and validation.
Related work is presented in Section \ref{sec:related}.
Conclusions and further work are provided in Section \ref{sec:conclusion}.
Equilibrium Point Analysis - EPA applied to reducing the cost of  
solving a txn processing systems is presented in the Appendix.

%----------------------------------------------------------
\section{Transaction Processing with Predeclared Lock Requests}\label{sec:static}
\vspace{5mm}

The effect of granularity of locking on txn response time is investigated in Thomasian 1985 \cite{Thom85},
Txns are activated after acquiring all locks, 
while txns with lock conflicts with currently active txns are held in a queue
until requested locks are released by completed txns.
Txns are processed in FCFS order.

Txn response time is the sum of queueing due to acquire all locks
and txn execution time at the computer system, 
which is represented by a product-form QN model.
We consider $J=5$ txn classes and a maximum degree of concurrency $K=2$,
since only txns in class ${\cal C}_1$ and ${\cal C}_2$ can be processed concurrently.
Txns in ${\cal C}_j$ are a fraction $f_j$ in Poisson arrival stream. 

Using the hierarchical decomposition method these throughputs
are then incorporated into the higher-level model which is a 2-dimensional CTMC.
One dimension is the composition of timesharing requests in execution
$${\cal S}_j , 1 \leq j \leq J \mbox{  and  }{\cal S}_{1,2}.$$
and another dimension the number of requests in the system.

Wallace and Rosenberg's 1966 {\it Recursive Queueing Analyzer - RQA} \cite{WaRo66} 
was used to succinctly specify the sparse regularly structured state transition matrix ({\bf $Q$}).
The number of states in the second dimension is set to be sufficiently large 
so that the fraction of txns lost due to the finite capacity is negligibly small for the given arrival rate.
An iterative method of the form
$$\underline{\pi} (k+1) = \underline{\pi}(k) (c {\bf Q} +{\bf I}),$$
where $c$ is a constant and ${\bf I}$ is the unity matrix 
Kleinrock 1975 \cite{Klei75}, Bolch et al. 2006 \cite{BGdT06}.

Given the state probabilities we can obtain the mean number of txns 
in different classes $\overline{N}_j, 1 \leq j \leq J$.
The mean txn response times follow as $R_j = \overline{N}_j / \lambda_j$.

The analysis can be extended to FCFS with skipping for static locking
as in the analysis of static locking in Thomasian and Ryu 1983 \cite{ThRy83} 
The latter analysis postulates a fine granularity of locking
and that lock requests are uniformly distributed over database granules. 

%-----------------------------------------------------------------------
\subsection*{Aggregating Multiple Transaction Classes}

The resulting system can be specified as:
$$T(k), 1 \leq k \leq K_{max}\mbox{  and  }T(k)=T(K_{max}) , k \geq K_{max}$$
and then incorporating the throughout in a higher birth-death model.

Txn throughput with a single class is a weighted sum according to txn frequencies.
Note that txns in the same class are not compatible with each other
and only txns in ${\cal C}_1$ and ${\cal C}_2$ can be executed together.
With at most two txns in execution and an infinite backlog of txns processed in FCFS order
we have the transition rate matrix $T$ among the execution states:
$$T_{j,j} = - \sum_{i \neq j} T_{j,i}, j=1,6.$$
Solving the set of linear equations yields the state probabilities.
$$\underline{p} {\bf T(2)} = 0 \mbox{ and  }\sum_{\forall(i)} p_i =1 .$$
The matrix for a closed task system with two tasks and 
compatible {\cal C}$_1$ and {\cal C}$_2$ classes is as follows.

\begin{eqnarray}\nonumber
\begin{tiny}
\begin{pmatrix}
T_{1,1} & 0 & \frac{f_3}{1-f_2}T_1  & \frac{f_4}{1- f_2} T_1
& \frac{f_5}{1 - f_2 } T_1 & \frac{f_1 f_2}{1-f_2} T_1  \\
0 & T_{2,2} & \frac{f_3} {1- f_1} T_2 &
\frac{f_4}{1-f_1} T_2 & \frac{f_5}{1 - f_1} T_2 & \frac {f_1 f_2}{1 - f_1} T_2 \\
f_1 (1-f_2) T_3 & f_2 (1- f_1) T_3 & T_{3.3} & f_4 T_3 & f_5 T_3 & 2 f_1 f_4 T_3       \\
f_1 (1- f_2) T_4 & f_2 (1 - f_1 ) T_4 & f_3 T_4 & T_{4,4} & f_5 T_4 & 2 f_1 f_2 T_4 \\
f_1 (1 - f_2) T_5 & f_2 (1 - f_1) T_5 & f_3 T_5 & f_4 T_5 & T_{5,5} & 2 f_1 f_2 T_5   \\
(1 -f_2) {T'}_2 & (1-f_1) {T'}_1 & 0 & 0 & 0 & T_{6,6}
\end{pmatrix}
\end{tiny}
\end{eqnarray}

After solving the set of linear equations to obtain the state probabilities $E_i, \forall{i}$
we can obtain the throughputs for class ${\cal C}_j$ with $k$ txns in execution.

\begin{eqnarray}
T_j (k) = \sum_{\forall{i}} P[E_i] T_j (E_i)
\end{eqnarray}

The overall txn throughput is

\begin{eqnarray}
T(k) = \sum_{\forall{j}} T_j (k)
\end{eqnarray}
The mean number of txns in {\cal C}$_j$ with $j$ txns in the systems is:

\begin{eqnarray}
\bar{N}_j (k) = \sum_{\forall{i}} P[E_i] | E_i |_j
\end{eqnarray}
where $|E_i|_j$ is the number of txns in class ${\cal C}_j$ executing in that state (zero or one).

Txn throughputs for the five classes with $k=2$ txns are:

$$ T_j (2) = P ({\cal S}_1) T_1 ({\cal S}_1)  + P ({\cal S}_{1,2}) {T}_j ({\cal  S}_{1,2}), \hspace{2mm} j=1,2. $$

$$ T_j (2) = P ({S_j}) T_j ({\cal S}_j) , \hspace{2mm} j=3,5. $$

For execution states with $k$ txns in the system are:
$$T_i (k)  / T_j (k) = f_i / f_j\mbox{ and }T_i (k) = f_i T(k),$$
where $T(k) = \sum_{\forall{i}} T_i (k)$.
For degree of concurrency $k=2$ the mean number of txns in different classes is:

\[ 
\overline{N}_1 (2) = P[{\cal S}_1] (1 + \frac{f_1}{1-f_2} )+ P( {\cal S}_{1,2} ) + f_1 \sum_{j=3}^5 P ({\cal S}_j) 
\]

\[ 
\overline{N}_2 (2) = P[{\cal S}_2] (1 + \frac{f_2}{1-f_1} )+ P({\cal S}_{1,2} ) + f_2 \sum_{j=3}^5 P ({\cal S}_j) 
\]

\begin{eqnarray}\nonumber
\overline{N}_j (2)=  
P[{\cal S_j}] (1+f_j) +
P[{\cal S}_1] [f_1 / (1-f_2) ]                          \\ 
\nonumber
P[{\cal S}_2] [f_j / (1-f_1) ], \hspace{2mm}3 \leq j \leq 5.
\end{eqnarray}

%The mean number of txns in the system is determined by solving the higher level model,
%which is a birth-death process to obtain $P_k$.

%The maximum degree of txn concurrency may not be attainable 
%due to lock conflicts for a coarse locking granularity,

%%%%%%%%%%%%%%%%%%%%%%%%%%%%%%%%%%%%%%%%%%%%%%%%%%%%%%%%%%%%%%%%%%%%%%%%%%%%%%
\section{Makespan of Task System with Multiprogrammed Tasks}\label{sec:tasksystem}
\vspace{5mm}

Given a task system is specified by a dag with precedence relationships among tasks
Tasks in Coffman and Denning 1973 \cite{CoDe73} have fixed execution times.
We consider a task system whose tasks are specified by their service demands 
at the devices of a multiprogrammed computer system.
In Thomasian and Bay 1986 we develop a hierarchical analysis 
to determine the makespan, the completion time of the task system. 

We consider a simple task system ${\cal \bf T}$ with six tasks.
Two complementary tasks are added: $\tau_0$, which precedes all tasks 
and $\tau_\infty$ which succeeds tasks with no successors otherwise.   
These two tasks are processed instantaneously.
$$ {\cal \bf T } = \{\tau_0,\tau_1,\tau_2,\tau_3,\tau_4,\tau_5,\tau_6,\tau_\infty\} $$
with the following precedence relationships:             \newline
$\tau_0 \prec \{ \tau_1, \tau_2 \} \prec \tau_3$,        \newline
$\tau_0 \prec \tau_4 \prec \{ \tau_5, \tau_6 \}$,        \newline
$\{\tau_3, \tau_5, \tau_6\} \prec \tau_\infty$.                                                      

The task system makespan is $C=\mbox{Init}_\infty$.
We are also interested in the initiation ($Init_i$), completion ($Comp_i$)
and execution $Exec_i=Comp_i - Init_i$ time of the $i^{th}$ task.
The execution time of a task is the time the task spends in the system. 

The task system ${\cal \bf T}$ leads to the CTMC for task execution states given in Table \ref{tab:states}.
Task combinations executed together known as tasksets are given in  a list.
An implicit instant transition from state $\{\tau_\infty\}$ to state $\{\tau_0\}$
can be postulated so the execution of the task system is repeated. 

\begin{table*}[ht]
\begin{small}
\begin{center}
\begin{tabular}{|c|c|c|c|c|c|c|c|}\hline
%----------------------------------------------------------------------
$L_0$ &$\{\tau_0\}$ &&&&&&                                             \\ \hline
%-----------------------------------------------------------------------
$L_1$ &$\{\tau_1,\tau_2,\tau_4\}$ &&&&&&                            \\ \hline                                 
%-----------------------------------------------------
$L_2$ &$\{\tau_1,\tau_4\}$, &$\{\tau_2,\tau_4\}$ &$\{\tau_1,\tau_2,\tau_5,\tau_6\}$ &&&& \\ \hline
%--------------------------------------------------------
$L_3$ &$\{ \tau_1,\tau_2,\tau_5\}$ & $\{\tau_1,\tau_2,\tau_6\}$ &
$\{ \tau_1,\tau_5,\tau_6\}$ &$\{\tau_2,\tau_5,\tau_6\}$ &$\{ \tau_3,\tau_4 \}$ && \\ \hline      
%----------------------------------------------------------------
$L_4$ &$\{ \tau_1,\tau_2\}$ & $\{\tau_1,\tau_5\}$  &$\{ \tau_1,\tau_6 \}$ 
&$\{\tau_2,\tau_5\}$ &$\{\tau_2,\tau_6\}$ &$\{ \tau_3, \tau_5, \tau_6\}$ &$\{ \tau_4\}$ \\ \hline
%-----------------------------------------------------------------------------------
$L_5$  &$\{\tau_1\}$ &$\{\tau_2\}$  &$\{\tau_3,T_5\}$ &$\{\tau_3,\tau_6\}$  &$\{\tau_5,\tau_6\}$ && \\ \hline  
%------------------------------------------------------------------------
$L_6$ &$\{\tau_3\}$ &$\{\tau_5\}$ &$\{\tau_6\}$  &&&& \\ \hline
%---------------------------------------------------------------------
$L_7$ &\{$\tau_\infty\}$ &&&&&&                         \\ \hline
\end{tabular}
\end{center}
\end{small}
\caption{\label{tab:states} CTMC has been built taking into account precedence relationships top to bottom.
Given that one task completes per level the number of levels equals the number of tasks.
Tasksets at lower levels are either a subset of tasksets at the higher level
or additional tasks being activated when precedence relationships are satisfied.}
\end{table*}

The completion of $\tau_4$ leads to the following transition
$$S=\{\tau_1,\tau_2,\tau_4\} \rightarrow \{ \tau_1, \tau_2, \tau_5, \tau_6 \}$$
The state holding time is the inverse of the sum of task throughputs, 
which determine the rates of an exponential distribution 
according  to the decomposition principle Courtois 1975 \cite{Cour75},
which is discussed informally in Lazowska et al. 1984 \cite{LZGS84}.
The notation used in this section is as follows:

\noindent
$I$: Number of tasks including two dummy tasks, which complete instantaneously.       \newline  
$L$: Number of CTMC levels with $L=I$, since one task completed  per level.           \newline
$S$: State representation.                                                            \newline
$S_\ell$: Set of states at level $\ell$.                                              \newline
$|S|=\{\tau_i,\tau_j,\ldots\}$: Set of tasks in execution at state ${\cal S}$.        \newline
${S}_i$: Set of states at which task $\tau_i$ is executed.                            \newline                 
${S}^+$: Immediate successors to state $S$.                                           \newline 
${S^-}$: Immediate predecessors to state $S$.                                          \newline
$P(|S|)$: Steady state probability of being in state $S$.                             \newline
$T_i(S)$: Completion rate or throughput of task $\tau_i \in |S|$.                     \newline
$T(S) = \sum_{\tau_i \in (|S|} T_i(S)$: Sum of completion rates at ${S}$.             \newline
$H(S)=[T(S)]^{-1}$: Mean holding time in $S$.                                         \newline                     
$b_R(S)$: Branching probability from $S$ to $R$.                                      \newline    
$p(R) = \sum_{S \in R_-} p(S) b_R(S)$ /* path probability to $R$ */                   \newline
/* The Mean delay to complete state $R$  is: */                                       \newline
$D(R)$ = $\sum_{S \in R_-} p(S) b_R(S) P(s) + H(R)$                                   \newline
$C$: = Completion time of all tasks.  

Path probability to reach state $R$.

\begin{eqnarray}\label{sec:pathprob}
p(S) = \sum_{R \in S^-} p(R) b_S(R).
\end{eqnarray}

The mean delay to the completion of state $R$ weighed by the path probabilities.

\begin{eqnarray}\label{sec:delay}
D(S) = H(S) + \sum_{R in S^+} p(R) b_S(R) D(R).
\end{eqnarray}

Initiation time of $\tau_i$ is a weighed sum of all delays for its activation in state $R$.

\begin{eqnarray}\label{eq:init}
\mbox{Init}_i = \sum_{ ( S \in R^- ) \land ( \tau_i \notin S ) \land ( \tau_i \in R ) } p(S)  D(S). 
\end{eqnarray}

The completion time of $\tau_i$ which completes at $S$ leads to $R$ which does not include $\tau_i$  

\begin{eqnarray}\label{eq:comp}
Comp_i = \sum_{ (S \in R^-) \land (\tau_i \in S) \land (\tau_i \notin R )} p(S) D(S).
\end{eqnarray}

Unnormalized state probabilities are computed level by level
by setting the probability of the initial state $S = \{ \tau_0  \}$ to one.

\begin{eqnarray}\label{eq:PR}
P(R) = \sum_{ S \in R^- } T(S) b_R(S) P(S). 
\end{eqnarray}
The state probabilities are normalized by 
$$\mbox{NormConstant}=\sum_{\forall S} P(S).$$

The execution time of $\tau_i$ is $Exec_i= Comp_i - Init_i$. 
Alternatively, completion time $C$ times the sum of state probabilities of states in which the task was executing.

\begin{eqnarray}
E_i = C \sum_{S\in {\cal S_i}} P(S).  
\end{eqnarray}

Each entry in the CTMC is represented as:
$$\left[ P(S); p({S}; D(S); T_i(S), \tau_i \in |S|; T(S), H(S),  b_R (S), R \in S^+ \right]$$

%%%%%%%%%%%%%%%%%%%%%%%%%%%%%%%%%%%%%%%%%%%%%
\begin{framed}
\subsection*{Procedure for Performance Analysis of Task System} 

Input: Set of $I$ tasks, precedence relationships and 
service demands at the $N$ devices of a multiprogrammed computer system: 
$$X_{i,n}, 1 \leq i \leq I, 1 \leq n \leq N.$$

Given $S=\{tau_0\}$ set $H(S)=0$, $P(S)=1$, $p(S)=1$ 

for levels $\ell=0$ to $L+1$ do 

for states $S \in {\cal S}_\ell$ do

Given that the completion rate of $\tau_i \in |S|$ is $T_i (S)$ 
$$T(S)=\sum_{\tau_i \in |{\cal S}} T_i (S)\mbox{  hence  }H(S)=1/T(S)$$

Determine all successor states to  ${\cal S}_\ell$  
and merge with the set of previously created states at $L_{\ell+1}$.
$${\cal R}_{\ell+1} = {\cal R}_{\ell+1} \cup R$$

Obtain probability of reaching state $R$ via $S$: $p(R) = p(S) \times  b_R(S)$ 

Completion of $\tau_i$ at $S$ leads to $R$ with probability: $b_R (S) = T_i(S)/T(S)$.

Path probability: $p(R) = p(R) + \sum_{S \in R^-} p(S) b_R(S)$
\vspace{-1mm}
$$D(R) = D(R) + \sum_{R \in S^-}  p_R(S) \times D(S)$$

Add to $\mbox{Init}_i$ tasks $\tau_i$ activated at this level using Eq. \ref{eq:init}.

Update the completion time of a task $\tau_I$ at this level using Eq. \ref{eq:comp}.

Obtain the steady state probability of $P(R)$ using Eq. (\ref{eq:PR}).

Update normalization constant for state probabilities $\mbox{Norm\_Constant} = \mbox{Norm\_Constant} + P(R)$

end   /* all tasks R in level $\ell$  */

end   /* level $L_\ell$ */

Normalize state probabilities  \newline
$P(S) = P(S)/ \mbox{Norm\_Constant }\hspace{2mm}\forall{S}$.

\end{framed}

Given the solution of the computer system model state probabilities can be used
to determine the mean device utilization when executing across all states.

Two numerical examples validated by simulation are provided in \cite{ThBa86}. 

%%%%%%%%%%%%%%%%%%%%%%%%%%%%%%%%%%%%%%%%%%%%%%%%%%%%%%%%%%%%%%%%%%%%%%%%%%%%%%%
\section{Simulation at Higher and Analysis at Lower Level}\label{sec:timesharing}
\vspace{5mm}

Hierarchical simulation is a more flexible method than building 
and solving a higher level CTMC for the analysis of task system performance.
It is computationally more expensive, 
since using the batch method the simulation has to be repeated 
to obtain confidence intervals at an acceptably high level Welch 1983 \cite{Welc83}. 

The method is specified in the context of performance analysis 
of a timesharing system with two sets of users generating requests Sauer \cite{Saue81}. 
The analysis is repeated in Thomasian and Gargeya 1984 \cite{ThGa84}. 

The first (resp. second) set of users are at $L_1$ (resp. $L_2$) terminals, 
which generate small class $C_1$ and large class $C_2$ requests. 
The think times at the terminals are exponentially distributed with means $Z_1$ and $Z_2$. 
The maximum MPL for processing $C_1$ and $C_2$ job classes are $M_1$ and $M_2$. 
The parameter settings used in experiments are based on Table 2 in Sauer 1981 \cite{Saue81},
which is repeated in Table \ref{tab:tab2}.

\begin{table}[ht]
\begin{small}
\begin{tabular}{|c|c|c|c|c|c|c|c|c|c|}\hline
Case  &1 &2 &3 &4 &5 &6 &7 &8 &9 \\ \hline \hline
%---------------------------------------------------
$L_1$ &20 &20 &20 &30 &30 &30 &40 &40 &40 \\ \hline 
$L_2$ &2 &2 &2 &3 &3 &3 &4 &4 &4 \\ \hline
$K_1$ &4 &3 &1 &7 &5 &2 &14 &9 &5 \\ \hline
$K_2$ &2 &1 &1 &2 &1 &1 &4 &3 &1 \\ \hline 
\end{tabular}
\end{small}
\caption{The nine cases considered in this study \label{tab:tab2}.}
\end{table}

Requests are processed at the CPU with the PS discipline 
and access four FCFS  disks with exponential service times uniform probabilities.
Think times and device service times are given in milliseconds as:
$$Z_1 = 5,000,   X_{CPU}^1=100,   X^1_{Disk_i}, 1 \leq i \leq 4 = 87.5$$
$$Z_2 = 100,000, X_{CPU}^2=2,000, X^2_{Disk_i}, 1 \leq i \leq 4 = 175$$

The closed QN, including the terminals, would be product-form if $M_j \geq L_j, j=1,2$, 
so that there would be no blocking due to MPL constraints.
A hierarchical solution method is required since realistically $M_j < L_j$,
which implies an extra delay in the memory queue.

The computer system is substituted with an FESC 
with the throughput characteristic for two classes:
$$T_j (k_1,k_2), \hspace{5mm} 0 \leq k_j \leq K_j, \hspace{5mm} j=1,2$$
There are $\prod_{j=1}^2 (N_j+1)$ states in the two-dimensional CTMC, 
since $0 \leq k_j \leq N_j,j=1,2$.
At state $(k_1,k_2)$ class $C_j$ job requests are issued at rate 
$$\Lambda_j = (L_j - k_j)/Z_j, j-1,2$$ 
and the throughputs obtained by solving the QN of the computer system are:
$$T_j (k_1,k_2), \hspace{2mm}  0 \leq k_j \leq K_j, \hspace{2mm} j=1,2.$$

The CTMC can also be build for an infinite source model, i.e.,  
by specifying the fraction of classes in the Poisson arrival stream. 
The number of states in the CTMC should be set to be sufficiently large,
so that probability of blocking due to finite capacity for the given arrival rate is negligibly small.
The set of linear equations to obtain CTMC's steady state probabilities  
can be solved using the Gauss-Seidel iterative method Stewart 2009 \cite{Stew09}.        
%\newline    %see e.g., \cite{Saue06}.
%\url{https://en.wikipedia.org/wiki/Gauss%E2%80%93Seidel_method}

%%%%%%%%%%%%%%%%%%%%%%%%%%%%%%%%%%%%%%%%%%%%%%%%%%%%%%%%%%%%%%%%%%%%%%%%%%%%%%%%%%%
\begin{framed} 
\subsection*{Procedure: Hierarchical Simulation of Timesharing System} 

\begin{description}

\item[1: Input simulation parameters.]                                                      

Number of classes of timesharing users: $J=2$.                                                \newline
Number of users or terminals in each class: $L_j, j=1,J$.                                     \newline
Maximum MPL in each class $K_j, j=1,J$.                                                       \newline
Number of active terminals/users in two classes $N_j = L_j, j=1,J$.                           \newline
Think times and job service demands at the CPU and four disks.                                \newline
Percent {\it Confidence Interval - CI} (say $\pm 5\%$)of job response times desired about the mean 
at a given {\it Confidence Level - CL} (say 95\%) using the batch means method \cite{Welc83}.  \newline
Settings such as: $NCompTarget_j=10,000,j=1,2$ $NumBatches=10$ need be adjusted to meet CL and CI target. 

\item[2: Solve lower level model.] 

Obtain throughputs by solving closed QN for all request compositions in two classes.
\vspace{-1mm}
$$T_j (k_1,k_2) \hspace{2mm}0 \leq k_j \leq K_j \hspace{2mm} j=1,2 $$

\item[3: Higher Level Discrete-Event Simulation.] 

\hspace{1mm}
(a) Initialization.            

$Clock=0$.       /* Simulation Clock */                  \newline
$BatchCtr=0$;    /* Batch means method */                \newline
for class $C_j j=1,2$ do                                 \newline
$count_j=0$    /* \#arrivals per class */                \newline
$N_j=L_j$      /* initialize \# of thinking users */     \newline
Sample $IntArvlTime$ from Exp($N_j/Z_j$)                 \newline
/* Set arrival time */                                   \newline
$ArvlTime_j = Clock + IntArvlTime$                       \newline                                 
/* Set departure times for all requests */               \newline
$DepartTime_j^k = \infty, 1 \leq k \leq K_j$             \newline
/* Sum class $C_j$ response times */                     \newline
$SumResp_j=0$                                            \newline
/* \# of completed jobs in $C_j$ */                      \newline
$NComp_j=0$                                              \newline
end do     /* do j */

\vspace{1mm}
(b) Scheduling the next event.

\begin{enumerate}
\item 
Determine most imminent event from ($N_1+N_2 + J \times K$) possibilities):     \newline
$Next = \mbox{min}[(ArvlTime_j,DepartTime_{j,k}$,                               \newline 
$1 \leq k \leq K_j, \hspace{5mm} j=1,2.$

If departure record request class and request's id: $jn=j$ and $kn=k$,             \newline                       
otherwise if arrival set arriving job class $jn$ based on the arrival stream.

\item 
Advance simulation time $Clock = Next$.                                        
\item 
If event an arrival goto (c),      \newline
else goto to (d).                
\end{enumerate}

(c) Arrival of a class ${\cal C}_{jn}$ request.

\begin{enumerate}
\item
$Count_{jn} = Count_{jn} + 1$;  
\item
/* one request generated at a time */
Sample from {\cal C}$_{jn}$ Exp($N_{jn}/Z_{jn}$). 
\item
/* next arrival time */ 
$ArvlTime_{jn} = Clock + IntArvlT_{jn}$.     \newline                      
\item
If $k_{jn} \geq K_{jn}$ enqueue              \newline
$Queue[Count_{jn}]=ArvlTime_{jn}$;           \newline
else goto (e)
\end{enumerate}

(d) Task completed at $Dept\_Time^{kn}_{jn}$.  

\begin{itemize}
\item $NComp_{jn}++$  /* increment completions */                                                               
\item $SumResp_{jn}=+ (Clock - Arrival_{kn,jn})$                       
\item if $[(NComp_1 \geq NCompTarget_1) \land$           \newline      
$ (NComp_2 \geq NCompTarget_2)]$ goto (g),  /* run  complete */              
\item $k_{jn}=k_{jn}-1$  /* degree of concurrency                                      
\item $N_{jn} = N_{jn}+1$.  /* active users */       
\item Obtain IntArvlT from Exp($N_{jn}/Z_{jn}$).     
\item $ArvlTime_{jn}=Clock+IntArvlT$.                
\item If $Queue[jn]$ is nonempty goto (e),          
else goto (b).                                       
\end{itemize}

(e) Activate an arriving or waiting task.                                      \newline
Activate $C_{jn}$ task, $k_{jn}++$                                             \newline
$Arrival_{kn,jn}  = TaskArvl[Count_{jn}]$                                      \newline
/* Set arrival time */                                                         \newline            
Sample execution time $E_{jn}^{kn}$ from Exp($T_{jn}(k_1,k_2)$).               \newline
$DepartTime_{jn}^{kn} = Clock + E_{jn,kn}$                                     \newline
goto (b) 

(g) BatchCtr++;                                                            \newline
$Resp_j^{BatchCtr} = \frac{SumResp_j}{NComp_j} , j=1,2$                          \newline
Exit if $BatchCtr++ \geq NumBatches$,                                            \newline
else goto (a)      

Batch means method to obtain mean response times and its CI at given CL was utilized.

\end{description}

\end{framed}
%%%%%%%%%%%%%%%%%%%%%%%%%%%%%%%%%%%%%%%%%%%%%%%%%%%%%%%%

%%%%%%%%%%%%%%%%%%%%%%%%%%%%%%%%%%%%%%%%%%%%%%%%%%%%%%%%%%%%%%%%%%%%%%%%%%%%%%%%%%%%%%%%%%%%%
\section{Effect of Transition Probabilities on Task Completion Times}\label{sec:distr}
\vspace{5mm}

We elaborates on Section 7.2 in Thomasian 2014 \cite{Thom14} that the distribution of the number 
of task cycles between CPU and disk processing affect the completion time of the task system with parallelism.

In Section 3.3.3 in Kobayashi 1978 \cite{Koba78} it is stated that routing in QNs 
need not be governed by a homogeneous first-order Markov chain and 
it is the mean number of visits to QN nodes that determines the usual performance metrics,
but this would affect task's sojourn time distribution.

The mean completion time of a 2-way F/J task system 
in a closed QN model is sensitive to the distribution of the number of cycles. 
A cyclic server model with two devices, a processor and a disk is postulated.
The disk is accessed following CPU processing and according to the cyclic server model 
after disk processing tasks may require additional CPU time or leave the system.
Note difference with CSM where jobs complete their processing at the CPU.
%This is not a closed model and a completed task leaves the system.

Consider the concurrent processing of two tasks whose number of cycles follows a geometric distribution:
$$p_n = (1-p) p^{n-1}, n \geq 1\mbox{ with a mean }\bar{n} = 1/(1-p).$$
The number of jobs completed in a time interval approaches a Poisson process 
when $p \rightarrow 1$, signifying a large number of cycles.
Poisson inter-departure times imply exponentially distributed residence times.
It can be shown that the geometrically distributed 
sum of exponential random variables is also exponentially distributed.
The argument based on thinning point processes in Salza and Lavenberg 1981 \cite{SaLa81}
does not require the per cycle residence times to be exponentially distributed or even i.i.d.

Consider the processing of two possibly heterogeneous tasks: $\tau_1$ and $\tau_2$,
which are processed concurrently at a computer system at state $S_{1,2}(\tau_1,\tau_2)$.
Given that $T_i(S_{1,2}$ is the completion rate of $\tau_i, i=1,2$ 
then the mean holding time for state $S_{1,2}$ is:
$$H(S_{1,2}) = [ T_1 (1,2) + T_2 (1,2)]^{-1}.$$
The completion of $\tau_i$ in $S_{1,2}$ leads to $S_j=\{\tau_{j \neq i}\}$
and the completion of $\tau_j$ leads to the completion of 2-way F/J task system.

Assuming that the transition rates are exponentially distributed,
the probability that $\tau_i$ completes first is:
$$p_i = T_i (S_{1,2})  H[S_{1,2}], i=1,2$$
The time to complete the two tasks is then:
$$C_{geo}  =  H(S_{1,2}) + p_1 H ( S_2 ) + p_2 H ( S_1 )).$$

Consider two tasks whose number of cycles is given as follows:
{\bf Case 1:} geometrically distributed with mean $\bar{n}=5$,
{\bf Case 2:} fixed with $n=5$.
The service demands per cycle is $\bar{x}_c=\bar{x}_d=\bar{x}=1$ at the CPU and the disk in both cases. 
%The two tasks with geometrically distributed processing times
%have an equal probability of finishing first.

Given balanced service demands task residence times for $N=K=2$ are: 
$$r(K)=(N+K-1)\bar{x}=3$$ 
according to Lazowska et al. \cite{LZGS84} and Thomasian 2023 \cite{Thom23}.
Due to the memoryless property of the geometric distribution 
the completion time for the two tasks in the two cases is:
$$C_{geo} = \bar{n} \times r(2) + \bar{n} \times r(1)  =  5 \times 3 + 5 \times 2  = 25.$$
$$C_{fixed} = n \times 3 =15.$$

We next consider the parallel processing of two tasks $\tau_1$ and $\tau_2$,
with a different number of geometric cycles which means $\bar{n}_1 = 5$ and $\bar{n}_2 =10$.
Setting the time per visit at the processor $\bar{x}_c=1$ and the disk $\bar{x}_d=1$
yields the service demands $X_C^1= X_D^1=5$ for $\tau_1$ and $X_C^2 = X_D^2 = 10$ for $\tau_2$.
Task throughputs when processed together can be obtained by solving the corresponding QN model.
$$T_1 (S_{1,2}) = 1/15\mbox{  and  }T_2 (S_{1,2}) = 1/30.$$
The mean holding time is then
$$H(S_{1,2}) = [ T_1 (S_{1,2}) + T_2 (S_{1,2}]^{-1} = 10.$$
The probability that $\tau_1$ or $\tau_2$ finishes first is given as:
$$
\begin{cases}
p_1 = T_1 (S_{1,2}) H(S_{1,2}) = (1/15) 10 = 2/3 \\
p_2 = T_2 (S_{1,2}) H(S_{1,2}) = (1/30) 10 = 1/3.
\end{cases}
$$
The mean residence time of these tasks is:
$$R(S_1)=X_c^1+X_d^1= 10\mbox{ and }R(S_2) = X_c^2 + X_d^2 =20.$$
It follows that the completion time for the geometric distribution is:
$$C_{geo}  = 10 + \frac{2}{3} (20) + \frac{1}{3} (10) \approx 26.77.$$

When the number of cycles is  fixed at $n_1=5$ and $n_2=10$,
then the two tasks share the CPU and disk for $\mbox{min}(n_1, n_2) = 5$ cycles.
The mean residence time per cycle for the balanced QN with $K=2$ task and $N=2$ devices 
based on a balanced F?J queueing system is \cite{LZGS84}, also see Thomasian 2023 \cite{Thom23}. 
$$r(N) = (K+N-1) \overline{x}= 3.$$
After $\tau_1$ completes, $\tau_2$ has five remaining cycles, where each cycle takes $r(1)=2$ time units.
The completion time is:
$$C_{fixed} = H(S_{1,2}) +  H(S_2) = n_1 \times 3 + (n_2-n_1) \times 2 =25.$$

In counting the number of remaining cycles we have in effect used the hybrid simulation method.
Given that the completion time of F/J and parallel task systems in general 
is sensitive to the number of cycles raises the issue that alternative methods 
to estimate the completion time of task systems.

%------------------------------------------------------
\section{Hybrid Simulation Method}\label{sec:hybrid}
\vspace{5mm}

Hybrid simulation is a hierarchical simulation method proposed by Schwetman 1978 \cite{Schw78},
which was applied to the CSM QN model with a single job class.
Tasks are specified by the number of required cycles and the loadings per cycle. 
The degree of concurrency is required for the analysis.

Tasks are specified by service demands per cycle and the initial number of required cycles. 
The remaining cycles for the $i^{th}$ task $\tau_i$ ($CR[i]$) is updated 
based on elapsed time $X=Next-Clock, Clock=Next$, where           
$$Next=\mbox{min}\{ ArvlTime, CT[i] \times CR[i]\}, 1 \leq i \leq n]$$
$CT[n]$ is the mean cycle time computed by solving the underlying QN model 
and $CR[i] = CR[i] - X/CT[i]$
As txns arrive and depart the degree of txn concurrency is updated as 
$n = n \pm 1$ and $CT[n]$ for active tasks is recomputed.

%This contradicts the central server model in \cite{Buze73} 
%where the number of cycles is geometrically distributed,
%but then the choice of this model was a matter of convenience.

The hybrid simulation method was extended to multiple job classes in Thomasian 1987 \cite{Thom87},
which deals with dynamic load balancing in a distributed system.
Simply stated it is better to process I/O bound jobs together with CPU bound jobs
and it is not the just remaining number of job cycles that matters.

The generalization of hybrid simulation proposed in Thomasian and Bay 1983 \cite{ThBa83}
does not require cyclic processing and can be applied to multiple job classes.
Rather than quantizing task processing times we may modify service demands according to Eq.~(\ref{eq:modified}), 
Given that $\tau_i$ starts its execution at device $d$ with loading $X_d^i$
and its mean residence time obtained by solving closed QN model for a given task composition is $R_i$, 
the residual job service demand after elapsed time $t$ is: 

\begin{eqnarray}\label{eq:modified}
\boxed{ X_d^i = X_d^i (1- t/R_i). } 
\end{eqnarray}

Tasks systems as in \cite{ThBa86} can be dealt with the hybrid simulation as follows.
Referring back to the example in Section \ref{sec:tasksystem} 
we refer to the processing of tasks $\tau_1,\tau_2,\tau_4$, 
which start processing concurrently.
The service demands at the CPU and Disk 1 and 2 for
$\tau_1,\tau_2,\tau_5,\tau_6$ are $(420,400,400)$
and for $\tau_3,\tau_4$ are $(620,600,600)$
and the number of cycles made by tasks are fixed.
Analysis of the QN model processing $\tau_1$, $\tau_2$, $\tau_4$ concurrently 
yields the mean residence time $R_1=R_2 < R_4$.
$\tau_4$'s residual loadings are obtained by multiplying by $(1-R_2/R_4$.

The modified hybrid simulation method is easier to implement 
and less costly than the method described in Section \ref{sec:tasksystem}.
In fact this method may be more accurate than the method based 
on decomposition that postulates exponentially distributed completion times.
This is especially so when the number of job cycles is not geometrically distributed.

With $n$ identical tasks when the distribution of number of cycles is uniformly distributed over $(0,x)$,
the minimum of the number of cycles to completion is: $r_{min}=x / {n+1}$

%%%%%%%%%%%%%%%%%%%%%%%%%%%%%%%%%%%%%%%%%%%%%%%%%%%%%%%%%%55
\begin{framed}

Batch jobs undergo different processing phases 
with different loadings from phase to phase, e.g., 
(i) loading and preprocessing of data, 
(ii) computation, 
(iii) visualization processing.
Rather dealing with a single set of task loading we assume loadings associated with 
its three phases designated as $\tau_1 \rightarrow \tau_2 \rightarrow \tau_3$ are known.
The residence time of the batch job is the sum of the residence times of three tasks.
The discussion is simplified by assuming that the computer also processes
transactions whose intensity remains fixed during the execution of the batch job.
Programs should be instrumented to signal change of phases for measurement purposes,
i.e., to report per phase loadings based on variation in resource consumption.
Phases accessing bottleneck resources will have increased residence times
while other phases will have shorter residence times.
Experiments are required if considering phases will yield significant differences
There is also the possibility of parallel processing via multitasking.

\end{framed}
%%%%%%%%%%%%%%%%%%%%%%%%%%%%%%%%%%%%%%%%%%%%%%%%%%%%%%%%%%%%%%%%%%%%%%%%%%%%%%%55

%Given elapsed time $t$ the interval can be adjusted to $r-t$.
%This approach is not applicable to the geometric distribution,
%because of its memoryless property \cite{Klei75}.
%Given a stream of incoming task requests sampling can be used 
%to determine the number the number of cycles from a given distribution.

%The $k^{th}$ order statistic for a uniform distribution has a $\beta$ distribution. 
%$$
%\beta(x) = \frac{x^{\alpha-1}(1-x)^{\beta-1}}
%{B(\alpha,\beta)}\mbox{  where }
%B((\alpha,\beta)=\frac{\Gamma(\alpha)\Gamma(\beta)}{\Gamma(\alpha + \beta)}
%$$
%where $G(x)=\int_0^\infty t^{x-1}e^{-t}dt$ is the Gamma function.

%------------------------------------------------------
\section{Related Work}\label{sec:related}
\vspace{5mm}

Given that many computations exhibit parallelism, 
several parallel models of computation have been developed, since the early days of computing. 
The early task system model based on a dag was described in Coffman and Denning 1973.

More complex relations in parallel processing can be represented by Petri nets Peterson 1981 \cite{Pete81}. 
The so-called UCLA graph model proposed in Martin 1966 \cite{Mart66} can be transformed into a Petri net.
but the proof of the equivalence of the two models by Kim Gostelow is flawed according to the book.

A major extension to Petri nets was the {\it Timed Petri Net - TPN} model 
by several researchers including Molloy 1982 \cite{Moll82}, 
which led to workshops on TPNs starting in 1985.
{\it Generalized Stochastic Petri Nets - GSPNs} allow immediate and 
exponentially distributed transitions Ajmone-Marsan et al. 1984 \cite{AMCB84}.
GSPNs are translatable to CTMCs which can the be solved using usual methods Stewart 2009 \cite{Stew09}. 
Further examples of this modeling approach are given in Ajmone-Marsan 1995 \cite{ACB+95}

{\it Research Queueing Package - RESQ} is a software tool for constructing and solving QN models. 
RESQ under a different name was imported to IBM Research 
from Univ. of Texas at Austin Sauer and MacNair 1983 \cite{SaMa83}.
It is also described in Sauer et al. 1977 \cite{SaRM77}.
RESQ provides a high level language for describing models also in hierarchical fashion.

Job execution in closed QN models allowing parallelism 
is considered in Heidelberger and Trivedi 1982 \cite{HeTr82}.
Jobs spawns two or more tasks at some point during execution, 
which execute independently of one another and do not require synchronization. 
An approximate solution method is developed and results 
of the approximation are compared to those of simulations. 
Bounds on the performance improvement due to overlap are derived.

The same authors in 1983 \cite{HeTr83} consider parallel tasks, 
which wait at the end of their execution for all of their siblings to finish execution.
Two approximate solution methods are developed and compared with simulations. 
The approximations are computationally efficient and highly accurate. 
A single instance of a task systems executing in a multiprogrammed computer 
is considered Thomasian and Bay 1986 \cite{ThBa86}. 

Concurrency in parallel processing systems is the topic of Kung 1984 \cite{Kung84}.
Jobs are modeled as dags whose nodes represent separate tasks.
Four variations are considered:
(1) jobs available at time zero or Poisson arrivals.
(2) dags: fixed or random.
(3) task service times: constant or exponentially distributed.
(4) fixed or infinite number of jobs.
Algorithm 1 minimizes the expected time to complete all jobs,
while Algorithm 2 maximizes processor utilization.

An algorithm to search for module assignments 
and replications to reduce task response times is explored by Chu and Leung \cite{ChLe87}. 
The objective function is the sum of task response time 
and a delay penalty for the violations of thread response time requirements. 
%Good module allocations and replications, which minimize task response time and yet satisfy 
%the thread response time requirements are determined by the algorithm.
The PS queueing discipline is used in this study,
so that with Poisson arrivals the output stream is Poisson
and the nodes can be analyzed separately when there are no synchronization delays,

Chu et al. \cite{ChSL91} propose two submodels for estimating task response times 
in distributed systems with resource contention. 
The first submodel is an extended QN to obtain module response times, 
which is solved by a decomposition technique to reduce computational cost 
by 2-3 orders of magnitude with respect to a direct approach. 

The second submodel is a weighed control-flow graph model 
from which task response time can be obtained by aggregating module response time 
in accordance with precedence relationships. 
Task response times estimated by the analytic model compare closely with simulation results. 
The model can be used to study the tradeoffs among module assignments, scheduling policies, 
interprocessor communications, and resource contentions in distributed processing systems.

Response time is affected by interprocessor communications,
precedence relationships, module assignments, hardware resource and data resource contention,  
and processor scheduling policies.
A task response time model that considers all of these factors is proposed.
A Petri net is used to represent resource contention,
and the task control flow graph represents precedence relationships.
A QN with resource contention is used to estimate the response time of each module.
Module response time consists of delays at the processors and resource queues and
is estimated by approximating the extended QN as independent finite capacity QNs.
The module response time is mapped onto a control flow graph,
and task response time is obtained by aggregating the module response times
in accordance with their precedence relationship in the control flow graph.
The task response time derived from the analytical model were validated against simulation results.

A modeling methodology for evaluating the execution of parallel programs containing looping constructs 
by estimating the average execution time of such a program in a distributed, 
multicomputer environment is proposed in Kapelnikov et al. \cite{KaME89,KaME92}. 
A combination of QN analysis of graph models of program behavior is considered in these studies. 
Complex programs are first decomposed into program segments, which are analyzed independently.
Combined results produce an approximate solution for the whole program. 

Task graphs represent parallel programs with dags, 
which are specified as a 4-tuple \{ {\bf T,P,A,E} \} by Menasce and Barroso 1992 \cite{MeBa92}.

\begin{itemize}
\item 
${\cal \bf T} = \{ \tau_1, \tau_2, \dots \} $ is the set of tasks of a parallel program.
\item 
${\cal \bf P}$ is the precedence relationship among tasks.  \newline
${\tau_j}$ can be activated when all $(\tau_i \in {\cal \bf P}$ complete their execution.
\item
{\cal \bf A} is an allocation function for tasks to processors $\tau_i \rightarrow P_j$. 
\item
{\cal \bf E:} determines the execution time based on processor speeds.
Similarly to \cite{ThBa86} tasks are specified by the their service demands on a computer system.
The execution time depends on task mix and hence can be determined by solving the QN.
\end{itemize}

A CTMC based technique to obtain the execution time of a task graph 
in a multiprogrammed computer system based on Thomasian and Bay 1986 \cite{ThBa86} 
is reported by Menasce et al. in \cite{MSP+95}.
The use of the static processor assignment policy 
called {\it Largest Task First Minimum Finish Time - LTFMFT}
shows that it is very sensitive to the degree of heterogeneity of the architecture, 
and that it outperforms all other policies analyzed.

Three dynamic assignment disciplines are compared and it is shown that in heterogeneous environments,
the disciplines that perform better are those that consider the structure of the task graph 
and not only the service demands of the individual tasks. 
The performance of heterogeneous architectures is compared 
with cost-equivalent homogeneous ones taking into account different scheduling policies.
Static and dynamic processor assignment disciplines are compared in terms of performance.

%------------------------------------------------------------------------
\section{Conclusion and Further Work}\label{sec:conclusion}
\vspace{5mm}

Hierarchical modeling is a useful tool in developing approximate analyses 
when the system does not lend itself to a direct solution or to reduce the cost of analysis or simulation 
by replacing a detailed model of a computer system by an FESC. 

Several instances of hierarchic analysis are discussed in this paper
with the goal of reducing the solution cost.
An efficient solution of a CTMC for a task system and
a simulation of a timesharing system with two jobs classes with MPL constraints.
In both cases tasks are processed on a multiprogrammed computer system representable as a product form QN.

When the tasks of a subtask system execute at the devices of independent computer systems
the completion time of subtask system may be used to determine the overall completion time.
The data transmission delays among nodes is the ratio of message length and data transmission rate,
assuming queueing delays are negligible.

In addition to the mean completion time the variance of completion times is of interest.
With the assumption that the holding time in each state of the CTMC is exponentially distributed
the variance of completion time is the variance of first passage time from the initial to final state in the CTMC.
The variance of first passage times in discrete-time Markov chains is derived in Hunter 2006 \cite{Hunt06}.
%The accuracy of this assumption remains to be verified.

Rather than considering all tasks in the distributed computer system at once, 
as done in \cite{ThBa86}, the mean and variance of completion time of task 
subsystems can be determined separately.
In a distributed system can use separate task systems 
to determine mean and variance of completion time per system,
which can be used to determine overall completion time.

Rather than the six task system in Section \ref{sec:tasksystem}
consider two subtask systems with three tasks each:                    \newline
\{ $\tau_1, \tau_2, \tau_3 \}$ and $\{ \tau_4, \tau_5, \tau_6  \}$.  \newline
The two subtasks are executed independly at two identical computer systems.
The mean makespan of of the task system can be determined  
by obtaining the mean and variance of each subtask system. 
%Two moment approximations for maxima are given in Crow et  al. 2007 \cite{CrGW07}.
%Formulas for special cases when the coefficient of variation $c < 1$  are provided. 
%When $c>1$ a 3-moment approximation to the Hyperexponential-2 distribution 
%as given by Eq. (4.4) in the paper is suggested.

Assuming a normal distribution a formula for the expected value of the maximum 
for three such random variables is given in Dasgupta 2023 \cite{Dasg23}.
In the case of nine random variables a two step computation can be used
e.g., by computing the maximum of $X_{1:3}^{max},X_{4:6}^{max},X_{7:9}^{max}$. 

The expected value of the maximum of $n$ i.i.d. random variables
with mean $\mu_X$ and standard deviation $\sigma_X$ of the components
of an F/J request according to David and Nagaraja 2003 \cite{DaNa03} is given as:

\vspace{-2mm}
\begin{eqnarray}\label{eq:Xmax}
\overline{X}_n^{max} \approx \mu_X + \sigma_X G(n).
\end{eqnarray}

A simulation based method to substitute disks with 
a {\it Shortest Access Time First - SATF} scheduling method with an FESC is discussed in  \cite{Thom11}.
Given $p$ pending requests the service time is reduced according to $p^{1/5}$,
i.e., service time is halved for $p=32$ random requests.
This is an example of using simulation at the lower level
to develop an analytic formula for disk service time.

The method developed in \cite{ThBa86} is incorporated in the SHARPE
reliability and performance modeling package at Duke University. \newline
\url{https://trivedi.pratt.duke.edu/}

Hierarchical modeling in the context of reliability and availability engineering 
is discussed in Chapter 16 in Trivedi and Bobbio 2017 \cite{TrBo17}.

The discussion is applicable to queueing analysis of communication networks
and manufacturing systems Buzacott and Shanthikumar 1993 \cite{BuSh93}.

%For example, the following approximation for the expected value of the maximum of $n$ random variables:
%$\{ X_1, X_2, \dots, X_n \}$. 
%with mean $\mu_X$ and standard deviation $\sigma_X$ is given in \cite{DaNa03}:
%$$\bar{X}_n^{max} \approx \mu_X + \sigma_X G(n)\mbox  { where } G(n) \leq \frac {n-1}{\sqrt{2n-1}}.$$

%--------------------------------------------------------------------------
\section*{Appendix: Equilibrium Point Approximation}\label{sec:EPA}
\vspace{5mm}

A multilevel analysis method of dynamic locking
is used to determine txn response times in Ryu and Thomasian 1990 \cite{RyTh90}.
The analysis takes into account hardware and data resource contention,
which is due to lock contention.
Txns encountering a lock conflict are blocked and
those whose lock requests lead to a deadlock are aborted and restarted.
Realistically a txn which has the least resources should be aborted
as in the case of the {\it Wait-depth Limited - WDL} policy \cite{FrRT92}.
In realistic models the probability of deadlock is negligibly small.

Txns arrive according to a Poisson process.
The number of activated txns  ($V$) is restricted
by the maximum multiprogramming level $(W)$, i.e., $V=\mbox{min}(A,W)$,
where $A$ is the number of txns in the system.
Txns making successful lock requests continue their execution,
while txns making a conflicting lock requests are blocked,
so that $J$ txns are active and $V-J$ txns are blocked.

Txns causing a deadlock are aborted and restarted
so that the number of active txns remains the same.
In fact deadlocks are rare \cite{ThRy91}
and have a negligible effect on performance and
their effect was ignored in further studies Thomasian 19993 \cite{Thom93}.
Blocked txns are activated when a txn completes or is aborted releasing all of its locks.
The lock and hardware resource contention models are used to determine txn throughput $\mu(V), 1 \leq V \leq W$,
which can then be used in conjunction with the arrival process to determine mean txn response time.

To compute the effective system throughput $\mu(V)$
we need to compute the steady-state probabilities of a Markov chain $P_T(J), 1\leq, J \leq V$.
The transition from state $J$ to $I$ is determined by the events upon the  completion of a txn step.
The probabilities for these events are determined at the completion time of txn steps.
$$\mu(V) =\sum_{J \leq V} P_S (J|V) t(J)\mbox{ respect for }1 \leq V \leq W.$$

An alternative solution method obviates the need
to compute the state probabilities for $1 \leq J leq V$
and reduces the solution cost of lower levels by a factor of $\approx V/\mbox{log}_2(V)$.

$A(J)$ which is the mean of the difference $I-J$ at completion instants can be computed as follows:
\vspace{-2mm}
\begin{eqnarray}\label{eq:A1}
A(J) = \sum_{I=J-1}^V (I-J) P_{tran} (j,i|V)
\end{eqnarray}
Given that $\bar{J}= \sum_{J=1}^V J P_S (J|V)$ the systems is in equilibrium we have

\vspace{-2mm}
\begin{eqnarray}\label{eq:A2}
A(J)=0\mbox{  for }J =\bar{j}
\end{eqnarray}
$A(J)$ is positive (resp. negative) when $J < \bar{J}$ (resp. $J > \bar{J}$.

Eq. \ref{eq:A2} can be solved using the bisection method,
since A(J) is a monotonically decreasing function in J.
The number of iterations is bounded by $\mbox{ceil}(\mbox{log}_2 (V))$.

The interpretation of this relationship is that $\bar{J}$ is the system's balance point
such that the system tends to stay there \cite{Cour75}
These studies dealt with potential overload due to thrashing in overloaded in virtual memory system,
where the system throughput increases as the {\it MultiProgramming Level - MPL} is increased,
but drops beyond a certain MPL.
This phenomenon is explored in the context of {\it 2-Phase Locking - 2PL}
using a simple model as the degree of txn concurrency is increased in \cite{Thom93}.

The above analysis is motivated by {\it Equilibrium Point Analysis - EPA},
which was applied to the analysis of multiaccess protocols in Tasaka \cite{Tasa86}:
``EPA is a fluid-type approximation which is only applied to the steady state.
It assumes that the systems is always at an equilibrium point.
Therefore, EPA does not necessitate calculating state transition probabilities.
An equilibrium point can easily be obtained
by numerically solving a set of simultaneous nonlinear equations.''
%Two related publications are \cite{FuTa83,Tasa86}.

An application of EPA is illustrated by an example if Figure 20 in \cite{DeBu78},
where there are $M=18$ users with think time $M / Z= 0.9$ so that $Z=18/0.90 =20$
The arrival rate of requests to the computer system is $a(N)=(M-N)/Z$.
The mean number requests at the computer systems is given
by the intersection of the throughput characteristic $T(N)$ and $a(N)$,
but the calculation is simplified by using intersection of two graphs
to determine $N_{intersect} \approx \bar{N}$
Otherwise we have to solve the following set of equations for $(N)$ setting $p(0)=1$ noting that they add to one.
\vspace{-1mm}
$$[(M-N)/X] p(N) = t(N-1)p(N-1), 1 \leq N \leq M$$
such that:
$$p(0) = [1+ \sum_{N=1}^M p(N)]^{-1}.$$

\subsection*{Acknowledgements}
This paper is partially based on papers the author coauthored PhD student Paul Bay, 
most notably Thomasian and Bay \cite{ThBa86}.
The Appendix is based on Thomasian and Ryu \cite{RyTh90}. 

\vspace{5mm}

%------------------------------------------------------------------


\begin{thebibliography}{10}

\vspace{5mm}
\bibitem{AdCD74}
T. L. Adam, K. M. Chandy, and J. R. Dickson.
A comparison of list schedules for parallel processing systems. 
Commun. ACM 17, 12 (1974), 685-690.

\bibitem{AMCB84}
M. Ajmone Marsan, G. Conte, and G. Balbo:
A class of generalized stochastic Petri nets for the performance evaluation of multiprocessor systems.
ACM Trans. on Computer Systems 2, 2 (May 1984),  93-122.

\bibitem{ACB+95}
M. Ajmone Marsan, G. Conte, G. Balbo, S. Donatelli, and G. Franceschinis.
Modelling with Generalised Stochastic Petri Nets.
John-Wiley \& Sons, 1995.

\bibitem{BCMP75}
F. Baskett, K. M. Chandy, R. R. Muntz, and F. G. Palacios.
Open, closed, and mixed networks of queues with different classes of customers.
J. ACM 22, 2 (1975), 248-260.

\bibitem{BGdT06}
G. Bolch, S. Greiner, H. de Meer, and K. S. Trivedi.
Queueing Networks and Markov Chains: Modeling and
Performance Evaluation with Computer Science Applications, 2nd ed.
Wiley-Interscience, 2006.

\bibitem{BuSh93}
J. A. Buzacott and J. G. Shanthikumar. 
Stochastic Models of Manufacturing Systems. 
Prentice Hall, 1993. 

\bibitem{Buze73}
J. P. Buzen.
Computational algorithms for closed queueing networks with exponential servers.
Commun. ACM 16(9): 527-531 (1973).
%https://dl.nadji.org/doi/pdf/10.1145/362342.362345

\bibitem{Buz+78}
J. P. Buzen, R. P. Goldberg, A. M. Langer, E. S. Lentz, H. S. Schwenk, D. A. Sheetz, and  A. W. Shum.
BEST/1 - Design of a tool for computer system capacity planning.
In Proc. AFIPS National Computer Conf. - NCC 1978, 447-455.

\bibitem{Buze78}
J. P. Buzen.
A Queueing Network Model of MVS.
ACM Computing Survey 10(3): 319-331 (1978).
%https://dl.nadji.org/doi/pdf/10.1145/356733.356738

\bibitem{ChSa80}
K. M. Chandy and C. H. Sauer.
Computational algorithms for product form queueing networks. 
Commun. ACM 23, 10 (Oct. 1980), 573-583.
%https://dl.nadji.org/doi/pdf/10.1145/359015.359020

%\bibitem{ChNe81}
%K. M. Chandy and D. Neuse.
%Fast accurate heuristic algorithms for queueing network models of computing systems. 
%Performance Evaluation 1, 1 (1981)): 96 (1981)

\bibitem{ChNe82}
K. Mani Chandy and D. Neuse.
Linearizer: A heuristic algorithm for queueing network models of computing systems. 
Commun. ACM 25, 2 (Feb. 1982), 126-134.

\bibitem{ChLe87}
W. W. Chu and K. K. Leung,  
Module replication  and  assignment  for  real-time  distributed processing systems. 
Proc. IEEE 75, 5 (May 1987), pp. 547-562.

\bibitem{ChSL91}
W. W. Chu, C.  Sit, and K. K. Leung.
Estimating task response time for real-time distributed systems with resource contentions.
IEEE Trans. on Software Engineering 17(10): 1076-1092 (October 1991).

\bibitem{CoDe73}
E. G. Coffman Jr. and P. J. Denning.
Operating Systems Theory.
Prentice-Hall 1973.

\bibitem{Cour75}
P.-J. Courtois.
Decomposability, instabilities, and saturation in multiprogramming systems. 
Commun. ACM 18(7): 371-377 (1975

%\bibitem{CrGW07}
%C. S. Crow IV, D. Goldberg, and W. Whitt.  
%Two-moment approximations for maxima.
%Operations Research 55, 3 (May-June 2007), 532-548.   
%$http://www.columbia.edu/~ww2040/CrowGoldbergOR2007ec.pdf

\bibitem{Dasg23}
A. Dasgupta.
A formula for the expected value of the maximum of
three independent normals and a sparse high dimensional case. 
Statistics Dept. at Purdue Univ. downloaded 2023 \newline
\begin{scriptsize}
\url{https://www.stat.purdue.edu/~dasgupta/orderstat.pdf}
\end{scriptsize}

\bibitem{DaNa03}
H. A. David and H. N. Nagaraja.
%Herbert A. David and H.~N. Nagaraja 2003. 
Order Statistics, 3rd edition, 
Wiley-Interscience 2003.

\bibitem{DeBu78}
P. J. Denning and J. P. Buzen:
The operational analysis of queueing network models. 
ACM Computing Surveys 10, 3 (1978), 225-261.

%\bibitem{FAIM83] 
%G. Fayolle, R. Iasnogorodski, and I. Mitrani. 
%The distribution of sojourn times in a queueing network with overtaking: Reduction to a boundary problem. 
%In Performance ‘83. Eds. A. K. Agrawala and S. K. Tripathi. 
%North-Holland, Amsterdam: 477-486, 
%FIM-SojournTimes_1983.pdf

\bibitem{FeBu73}
E. B. Fernandez and B. Bussell.
Bounds on the number of processors and time for multiprocessor optimal schedules. 
IEEE Trans. Computers 22, 8 (1973), 745-751.

\bibitem{FrRT92}
P. A. Franaszek, J. T. Robinson, and A. Thomasian.
Concurrency control for high contention environments. 
ACM Trans. Database Systems 17, 2 (1992), 304-345.

%\bibitem{GeNT04}
%L. Georgiadis, C. Nikolaou, A. Thomasian:
%A Fair Workload Allocation Policy for heterogeneous systems. 
%J. Parallel Distributed Computing 64(4): 507-519 (2004)

%\bibitem{GoNe67}
%W. J. Gordon, G. F. Newell:
%Closed Queuing Systems with Exponential Systems.
%Operations Research 15(2): 254-267 (1967).

%\bibitem{GhKa91}
%M. Ghodsi and K. Kant.
%Performance Analysis of Parallel Search Algorithms on Multiprocessor Systems.
%Performance Evaluation 13(1): 67–81 (1991)

%\bibitem{Gij+06}
%B.M. M.Gijsen, R .D.van der Mei, P. Engelberts, J.L.van den Berga, K.M. C.van Wingerden.
%Sojourn Time Approximations in Queueing Networks with Feedback.
%Performance Evaluation 63(8): 743-758 (August 2006).

%\bibitem{HaPa93}
%P. G. Harrison,  N. M. Patel. 
%Performance Modeling of Communication Networks and Computer Architectures.
%Addison-Wesley, 1991.

\bibitem{HeTr82}
P. Heidelberger and K. S. Trivedi:
Queueing Network Models for Parallel Processing with Asynchronous Tasks.
IEEE Trans. Computers 31, 11 (Nov. 1982), 1099-1109.

\bibitem{HeTr83}
P. Heidelberger and K. S. Trivedi:
Analytic Queueing Models for Programs with Internal Concurrency.
IEEE Trans. Computers 32, 1 (Jan. 1983), 73-82.

\bibitem{Hunt06}
J. J. Hunter.
Variances of first passage times in a Markov Chain with applications to mixing times
Res. Lett. Inf. Math. Sci., 10 (2006), 17-48.
%\url{https://mro.massey.thna81.nz/bitstream/handle/10179/4484/Variances_of_first_passage_times_in_a_Markov_chain_with_applications_to_mixing_times.pdf}

\bibitem{Jack57}
J. R. Jackson. 
Networks of Waiting Lines.
Operations Research 5, 4 (1957), 516-521.

\bibitem{KaME89}
A. Kapelnikov, R. R. Muntz, and M. D. Ercegovac.
A Modeling Methodology for the Analysis of Concurrent Systems and Computations.
J. Parallel Distributed Computing 6, 3 (1989), 568-597.

\bibitem{KaME92}
A. Kapelnikov, R. R. Muntz, and M. D. Ercegovac.
A methodology for performance analysis of parallel computations with looping constructs. 
J. Parallel Distributed Computing 14, 2 (1992), 105-120.

%\bibitem{KrGr84}   
%A. Krezesinski and J. Greyling, 
%Improved linearizer methods for queueing networks with queue dependent centers. 
%In Proc. ACM SIGMETRICS Conf. on Measurement and Modeling of Computer Systems, 1984, 41-51.

\bibitem{Klei75}
L. Kleinrock.
Queueing Systems, Vol I: Theory.
Wiley-Interscience 1975.

\bibitem{Klei76}
L. Kleinrock.
Queueing Systems, Vol. II: Computer Applications,
Wiley-Interscience 1976.

\bibitem{Koba78}
H. Kobayashi.
System Design and Performance Analysis Using Analytic Models.
Chapter 3 in K. M. Chandy and R. T. Yeh.
Current Trends in Programming Methodology, Vol. III: Software Modeling,  
Prentice-Hall 1978, 72-114.

\bibitem{KoMa09}
H. Kobayashi and B. L. Mark.
System Modeling and Analysis:
Foundations of System Performance Evaluation.
Pearson, 2009.

\bibitem{Kung84}
K. C.-Y. Kung.
Concurrency in Parallel Processing Systems.
Ph.D. Dissertation. 
Computer Science Department, UCLA, 1984.
%https://www.osti.gov/bibitem/5338150-concurrency-parallel-processing-systems  %Kung84.pdf

\bibitem{Lave83}
S. S. Lavenberg.
Computer Performance Modeling Handbook.
Academic Press 1983.

\bibitem{LZGS84}
E. D. Lazowska, J. Zahorjan, G. Scott Graham, and K. C. Sevcik:
Quantitative System Performance: Computer System Analysis Using Queueing Network Models
Prentice-Hall 1984.

\bibitem{Mart66}
D. F. Martin.
The Automatic Assignment and Sequencing of Computations on Parallel Processor Systems.
Ph.D. Thesis, U. of California, Los Angeles, Jan. I966.

\bibitem{MeAl89}
D. A. Menasce and V. A. F. Almeida.
Analytic Models of Supercomputer Performance in Multiprogramming Environments.
Int'l J. High Performance Computing Applications 3 2 (1989), 71-91.

\bibitem{MeBa92}
D. A. Menasce and L. A. Barroso.
A methodology for performance evaluation of parallel applications on multiprocessors.
J. Parallel Distributed Computing - JPDC 14, 1 (1992), 1-14.

\bibitem{MSP+95}
D. A. Menasce, D. Saha, S. C.~S. Porto, V. Almeida, and S. K. Tripathi.
%Daniel A. Menasce, Debanjan Saha, Stella C. S. Porto, Virgilio Almeida, Satish K. Tripathi.
Static and dynamic processor scheduling disciplines in heterogeneous parallel architectures.
J. Parallel Distributed Computing - JPDC 28, 1 (Jan. 1995),  1-18.
%https://www.academia.edu/2723678/Static_and_dynamic_processor_scheduling_disciplines_in_heterogeneous_parallel_architectures

\bibitem{Moll82}
M. K. Molloy:
Performance analysis using stochastic Petri nets.
IEEE Trans. Computers 31(9): 913-917 (1982)

%\bibitem{PaKT90}
%K. R. Pattipati, M. M. Kostreva, and J. L. Teele.   
%Approximate mean value analysis algorithms for queuing networks: 
%Existence, uniqueness, and convergence results. 
%J. ACM 37, 3 (1990), 643-673.
%https://dl.acm.org/doi/pdf/10.1145/79147.214074

\bibitem{Pete81}
J. L. Peterson.
Petri Net Theory and the Modeling of Systems.
Prentice-Hall 1981.

%bibitem{Raat89}
%K. E. E. Raatikainen:
%Approximating response time distributions. 
%SIGMETRICS 1989: 190-199
%https://dl.nadji.org/doi/pdf/10.1145/75372.75393

%\bibitem{Raat92}
%Kimmo E. E. Raatikainen.
%Modeling service distributions in queueing network simulation. 
%Simulation 59(2): 116-126 (1992)

\bibitem{ReKo75}
M. Reiser and H. Kobayashi.
Queuing Networks with Multiple Closed Chains: Theory and Computational Algorithms.
IBM J. Research \& Development 19, 3 (1975), 283-294.

\bibitem{ReLa80}
M. Reiser and S. S. Lavenberg.
Mean-value Analysis of Closed Multi-chain Queuing Networks.
J. ACM 27, 2 (1980), 313-322.

\bibitem{Reis00}
M. Reiser:
Mean value analysis: A personal account. 
Performance Evaluation 2000, 491-504

\bibitem{RyTh90}
I. K. Ryu and A. Thomasian.
Analysis of database performance with dynamic locking.
J. ACM 37, 3 (1990), 491-523.

\bibitem{SaTP96}
R. A. Sahner, K. S. Trivedi, and A. Puliafito. 
Performance and Reliability Analysis of Computer Systems:
An Example-Based Approach Using the SHARPE Software Package. 
Kluwer 1996.

\bibitem{SaLa81}
S. Salza and S. S. Lavenberg.
Approximating response time distributions in closed queueing network models of computer performance.
In Proc. 8th Int'l Symp. on Compute Performance Modelling, Measurement and Evaluation, 1981, 
F. J. Kylstra, Ed., 133-144.

%\bibitem{Saue06}
%T. Sauer. 
%Numerical Analysis, 2nd ed. 
%Pearson Education 2006.
%\url{https://en.wikipedia.org/wiki/Gauss%E2%80%93Seidel_method}

\bibitem{SaCh75}
C. H. Sauer and K. M. Chandy.
Approximate analysis of central server models.
IBM J. Research \& Development 19, 3 (1975), 301-313.
%https://www.researchgate.net/publication/220498093_Approximate_Analysis_of_Central_Server_Models

\bibitem{SaRM77}
C. H. Sauer,  M. Reiser, and  E. A. MacNair.  
RESQ — A  package  for  solution  of  generalized  queueing  networks.  
In Proc. Nat'l Computer Conf. 1977, 978-986.

%\bibitem{SaMa78}
%C. H. Sauer and E. A. MacNair.
%Simultaneous resource possession in queueing models of computers.
%Research Report RC-6971, IBM T. J. Watson Research Center, Feb. 1978.

\bibitem{Saue81}
C. H. Sauer.
Approximate solution of queueing networks with simultaneous resource possession. 
IBM J. Research \& Development 25, 6 (Nov.-Dec. 1981), 894-903.
%https://www.researchgate.net/publication/224104090_Approximate_Solution_of_Queueing_Networks_with_Simultaneous_Resource_Possession

\bibitem{SaMa83}
C. H. Sauer and E. A. MacNair,
Extended Queueing Network Models.
Chapter 8 in Computer Performance Handbook, S. S. Lavenberg, (ed.), 1983. 

\bibitem{Schw78}
H. D. Schwetman.
Hybrid simulation models of computer systems. 
Commun. ACM 21, 9 (Sept. 1978), 718-723.
%https://dl.nadji.org/doi/pdf/10.1145/359588.359594

\bibitem{Stew09}
W. J. Stewart.
Probability, Markov Chains, Queues, and Simulation: 
The Mathematical Basis of Performance Modeling
Princeton Univ. Press. 2009.

%\bibitem{SDdT95}                             
%R. Suri, G. W. W. Diehl, S. de Treville, and M. J. Tomsicek.   
%%Rajan Suri, Gregory W. W. Diehl, Suzanne de Treville, Michael J. Tomsicek.
%From CAN-Q to MPX: Evolution of queuing software for manufacturing. 
%Interfaces 25, 5 (1995), 128-150

%\bibitem{SuSV07}
%R. Suri, S, Sahu, and M. Vernon.
%Approximate mean-value analysis for closed queueing networks with multiple-server stations.
%In Proc. 2007 Industrial Eng. Research Conf, 1-6.
%https://pages.cs.wisc.edu/~vernon/papers/poems.07ierc.pdf

%\bibitem{Taka91}
%H. Takagi.
%Queueing Analysis: Foundations of Performance Evaluation, Vol. 1:
%Vacation and Priority Systems, Part 1.
%North-Holland 1991.

%\bibitem{Tasa86a}
%S. Tasaka.
%Dynamic behavior of a CSMA/CD system with population of buffered users.
%IEEE Transactions on Communications COM-34 (1986). 576-586.

\bibitem{Tasa86}
S. Tasaka.
Performance Analysis of Multiple Access Protocols.
The MIT Press 1986.
%https://www.google.com/books/edition/Performance_Analysis_of_Multiple_Access/PMRY3J0a5UYC

\bibitem{ThNa81} 
A. Thomasian and B. Nadji.
Algorithms for queueing network models of multiprogrammed computer systems.
Computer Performance 2, 3 (Sept. 1981), 100-123.

\bibitem{ThRy83}
A. Thomasian and I. K. Ryu.
A decomposition solution to the queueing network model of 
the centralized DBMS with static locking. 
In Proc. ACM SIGMETRICS on Measurement and Modeling of Computer Systems - SIGMERTICS 1983, 82-92.

\bibitem{ThBa83}
A. Thomasian and P. F. Bay.
Queueing network models for parallel processing of task systems. 
In Proc. Int'l Conf. on Parallel Processing - ICPP 1983, 421-428

\bibitem{ThGa84}
A. Thomasian and K. Gargeya.
Speeding up computer system simulations using hierarchical modeling. 
ACM SIGMETRICS Perform. Evaluation Review 12, 4 (1984), 34-39.

%\bibitem{ThBa85}
%A. Thomasian and P. F. Bay.
%Performance analysis of task systems using a queueing network model. 
%In Proc. Int'l Workshop on Timed Petri Nets - PNPM 1985: 234-242.

\bibitem{Thom85}
A. Thomasian.
Performance evaluation of centralized databases with static locking.
IEEE Trans. Software Eng. TSE-11, 4 (April 1985), 346-355.

\bibitem{ThBa86}
A. Thomasian and P. F. Bay.
Analytic queueing network models for parallel processing of task systems.
IEEE Trans. Computers 35, 12 (Dec. 1986), 1045-1054.

\bibitem{Thom87}
A. Thomasian.
A performance study of dynamic load balancing in distributed systems. 
In Proc. Int'l Conf. on Distributed Computing Systems - ICDCS 1987: 178-184

\bibitem{ThRy91}
A. Thomasian and I. K. Ryu.
Performance analysis of two-phase locking. 
IEEE Trans. Software Eng. 17, 5 (May 1991), 386-402.

\bibitem{Thom93}
A. Thomasian.
Two-phase locking performance and its thrashing behavior. 
ACM Trans. Database Systems 18, 4 (1993), 579-625.

\bibitem{Thom11}
A. Thomasian.
Survey and analysis of disk scheduling methods. 
ACM SIGARCH Computer Architecture Newsletter 39, 2 (2011), 8-25.

\bibitem{Thom14}
A. Thomasian:
Analysis of fork/join and related queueing systems.
ACM Computing Surveys 47, 2 (Aug. 2014), 17:1-17:71.

\bibitem{Thom23}
A. Thomasian
Unbalanced job approximation using Taylor series expansion and review of performance bounds.
\url{https://doi.org/10.48550/arXiv.2309.15172}

\bibitem{Triv01}
K. S. Trivedi.
Probabilistic and Statistics with Reliability, Queueing and Computer Science Applications, 2nd ed.
Wiley 2001.

\bibitem{TrBo17}
K. S. Trivedi and A. Bobbio.
Reliability and Availability Engineering: Modeling, Analysis, and Applications.
Cambridge Univ. Press, 2017.

\bibitem{WaRo66}
V. L. Wallace and R. S. Rosenberg.
Markovian models and numerical analysis of computer system behavior.
In Proc. AFIPS Spring Joint Computer Conf. - SJCC 1966, Vol. 27, 141-148.

\bibitem{Welc83}
P. D. Welch.
Statistical Analysis of Simulation Results.
Chapter 6 Computer Performance Handbook, S.S. Lavenberg (ed.), 1983.

\bibitem{WiBh76}
A. C. Williams and R. A. Bhandiwad.
A generating function approach to queueing network analysis of multiprogrammed computer systems.
Networks 6, 1 (1976), 1-22. 

%\bibitem{ZhDo16}
%L. Zhang and D. G. Down.      
%Lei Zhang, Douglas G. Down
%A stable mean value analysis algorithm for closed systems with load-dependent queues. 
%In Proc. 10th EAI Int'l Conf. on Performance Evaluation Methodologies and Tools, VALUETOOLS 2016
%Chapter04.pdf

%\bibitem{ZhDo19}
%L. Zhang and D. G. Down.           
%SMVA: A stable mean value analysis algorithm for closed systems with load-dependent queues
%Systems Modeling: Methodologies and Tools, 2019.	

\end{thebibliography}
\end{document}